# AusClinBench: a deterministically scored benchmark of large language models against current Australian clinical policy

**Daniel S Epstein**[1,2] MBBS(Hons) BMedSc(Hons) FRACGP PhD ORCID: 0000-0003-2215-7749

1. Department of General Practice, Monash University, Melbourne, Victoria, Australia
2. The Long Game Project, Melbourne, Victoria, Australia

Correspondence: dan.epstein@monash.edu

## Abstract

**Background.** AI models are already used in clinical practice, and they are here to stay. They inform clinical decisions and are currently writing a lot of the future medical history records for our patients. In Australian clinical software they are an unseen ingredient: the practitioner using the product is rarely told which model it runs on. Existing medical benchmarks for AI models are written for other jurisdictions. None measures whether a model knows the current Australian rules that a clinician must follow.

**Methods.** We wrote 142 questions from Australian clinical guidelines and schedules (the PBS Schedule, MBS Online, the Australian Immunisation Handbook, Ahpra guidance and national guidelines) across five domains: PBS authority, MBS billing, Ahpra obligations, triage red flags, and immunisation and screening. Each question is a short vignette with one verifiable answer, scored by exact match against a key. We asked each question under two conditions: blind (the question alone) and rule-supplied (the same question plus the governing clause, the text of the rule the key comes from). A question answered wrongly blind and correctly with the rule supplied was a recall failure. One answered wrongly under both was a reasoning failure. We tested 28 AI models between July and October 2026.

**Results.** Blind accuracy ranged from 51.5% to 92.7%. The highest-scoring models could not be separated from one another, by interval overlap or by paired tests. Each model answered each question in both conditions, giving a model-question pair. Of 791 failed pairs, 684 (86%, bootstrap interval 78 to 94) were recall failures, 81 were reasoning failures, and 26 were right blind but wrong with the rule. Confident-error rates (a wrong answer given without saying they 'don't know') ranged from 6.6% to 47.8%. On five immunisation questions where the guideline gives no directive, 82% of answers stated one anyway. Accuracy was lowest in PBS authority (56.0%), below three of the other four domains. On 20 Aboriginal and Torres Strait Islander health questions, models did no worse overall, but on a few of them the wrong answers gave the rule for the general population, an older rule or another country's rule.

**Conclusions.** Models differ widely in how well they know the Australian rules a clinician must follow, and most of their errors disappear when the rule is supplied. We propose that benchmarks like this can give regulators a practical safety measure: the model is an ingredient of a clinical product, and a product may use any model that meets a set standard on local rules. Such a standard would sit beside product evaluation, not replace it. The answer keys are one general practitioner's and have not yet been independently reviewed. The benchmark measures the base models. It does not evaluate clinical products or the software around a model, and it does not license any model for clinical use.

## 1. Introduction

Australian clinical practice runs on rules that change by legislation and by schedule: which medicines the Pharmaceutical Benefits Scheme (PBS) subsidises and on what conditions, which Medicare Benefits Schedule (MBS) items a doctor may bill and in what combination, which vaccines are funded for which groups from what age, and what the medical regulator (Ahpra) requires. A model that gives a confident, out-of-date answer to one of these questions should be considered less safe than a model that is correct, or admits it does not know. It can cost a practice money, deny a patient a subsidised script, or expose a clinician to a compliance finding.

AI models now sit inside transcription tools, inbox assistants and decision-support products used in Australian healthcare practice. The vendor chooses the model, and in the author's experience as a practising general practitioner the practice is rarely told which model is running under the hood. We know of no public measurement of how well these models know the Australian rules. Testing models on local rules can show where they fall back on another country's rule, or on the general rule for a group the local rules treat differently. It can also show which models suit the conditions where they are used. In Australia that includes Aboriginal and Torres Strait Islander people, for whom many funding, vaccination and treatment rules differ from those for the general population.

Published medical benchmarks do not fill this gap. The widely used ones test diagnosis and management knowledge drawn from medical licensing examinations, and are written for the United States or for no jurisdiction in particular [6,7,8]. Evaluation practice has also moved towards grading model output with other AI models [9], which introduces a grader whose own failure modes are rarely reported. Benchmarks of this era have been criticised for an unstated construct, contamination, and grading that is not itself validated by practising clinicians [10].

AusClinBench measures the Australian policy layer, with deterministic scoring and no model in the grading loop. Two design choices shape the rest of the method. First, it tests policy, not diagnosis. Policy changes yearly, so the items are refreshed rather than left to go out of date, and each key has to be checked against the current edition of its source. One key in this release was not, and section 3.8 reports and corrects it. Second, every item runs blind and with the governing clause supplied, so a failure can be classed as a recall problem or a reasoning problem. The two have different fixes.

The paper asks one primary question, fixed before the first run: is the leading model's accuracy on these rules measurably below, or above, a set threshold (section 2.9)? It then asks what the results mean for regulation, and whether a benchmark like this could serve as a safety measure for the models inside clinical products (section 4).

## 2. Methods

### 2.1 Item construction and provenance

Each item is a short clinical vignette, one question with a single verifiable answer, and an answer key written directly from a primary source. Version 1.4 has 142 items across five domains (Table 1). Items were written by one practising Australian general practitioner from the PBS Schedule [2], MBS Online [3], the Australian Immunisation Handbook [4], Ahpra and Medical Board guidance [5], and national clinical guidelines (for example the Royal Children's Hospital clinical practice guidelines). The adult and paediatric mix of the triage items was not recorded.

| Domain | Items | Public | What it tests |
|---|---|---|---|
| MBS billing | 52 | 12 | Item eligibility, time tiers, co-claiming restrictions, health assessments |
| Immunisation and screening | 33 | 5 | Schedule ages, dose intervals, screening intervals, funded groups |
| PBS authority | 24 | 5 | Authority and streamlined-authority requirements, initiation criteria |
| Triage red flags | 20 | 4 | Escalation thresholds and first-aid rules |
| Ahpra obligations | 13 | 4 | Regulatory and prescribing obligations under the National Law |
| **Total** | **142** | **30** | |

*Table 1. Items by domain. "Public" is the published sample. The remaining 112 items are held out (section 2.8). Ten recall-only items and five equipoise items are excluded from the failure decomposition (section 2.6).*

Each item records, when it is written, the source URL, the access date, a verbatim quote of the sentence that grounds the key, and an effective-from date. An automated freshness check re-fetches each source and re-verifies the quote.

Items are synthetic: they are written from the public rule, never from a patient record. A provenance guard in the test suite fails the build if any item declares another derivation.

Ten items are tagged recall-only: the key is a fact (an MBS item number, for instance) that no quoted clause can entail without stating it, so the rule-supplied condition is uninformative for them. Five items are equipoise items, where the guideline itself declines to give a directive and the correct answer is to say so. Recall-only items count on the blind board. Equipoise items are scored separately, each paired with a settled item on the same topic, and are not on the board. Both groups are excluded from the failure decomposition (section 2.6).

### 2.2 Conditions

Every item runs twice per model.

- **Blind.** The vignette and question only. This is the deployed-without-retrieval condition: what the model asserts when asked mid-consult.
- **Rule-supplied.** The same, plus the cited governing clause. This removes recall and tests whether the model can apply the rule it is handed. The clause is the exact one the key was written from, so this condition is retrieval at its best, not retrieval as a product would do it.

Throughout, a *cell* is one model answering one item under one condition, and a *pair* is one model's two cells for the same item. The *board* is the full set of results in Table 2: 28 models, 142 items, 7,952 cells.

A third condition, blind with web search enabled, ran for the ten July models only and is reported in section 3.11.

### 2.3 Models and serving routes

Twenty-eight models were tested (Table 2, Appendix A). Ten were run on 12 and 13 July 2026, fifteen on 6 and 7 September 2026, and three on 9 and 13 October 2026. The July and September calls went through one API layer (OpenRouter) with identical prompts, temperature 0, a fixed JSON response schema, provider fallbacks disabled and provider-side data collection denied. No reasoning or effort setting was sent, so each model ran at its provider's default. The API layer records which provider served each call, and some open-weight models were served by more than one provider across runs (Appendix A). The quantisation each provider used for open-weight models was not recorded. The model slugs were verified against the live catalogue on the day of each run. A newer model is added to the board, never substituted for its predecessor. Claude Fable 5 and Claude Fable 5.1 are different models and have separate rows.

The three October models (Claude Sonnet 5.5 and Claude Haiku 5.5 on 9 October, Claude Opus 5.5 on 13 October) ran through a Claude Code subscription, not through the API layer. The question text and system prompt were identical. The response format was added to the prompt as an instruction, because this route cannot enforce a schema. No temperature was sent, and the route offers no per-call data-collection control. The harness sent its system prompt as the request's system prompt, as on the API layer. Whether the route added a prompt of its own, and its tool access and thinking setting, are outside the harness and were not recorded, so these three rows differ from the rest in route, prompting, sampling and date at once. They are shown in a separate block of Table 2 and starred in Figures 1 and 2, and their scores should be read as a separate cohort (limitation 9). Section 3.4 gives the main pooled result without them. No earlier row was re-run, so no earlier score moved.

Two September models could not be measured and are excluded, not scored: Gemma 4 31B (every one of 284 calls rejected by the provider, HTTP 405) and Qwen 3.8 27B (67 of 284 replies empty, so any score would rest on a biased subset of items).

### 2.4 Scoring and outcomes

Models answer in a fixed structured form with named fields. A response is correct only when every field matches the key exactly. There is no partial credit and no model grades another model. The form includes an explicit way to decline. Every answer lands in one of three buckets: correct, declined (never scored correct, reported separately), or confident error (wrong, with no decline). Blind accuracy is the primary measure, because the stopping rule uses it. The confident-error rate is the main safety measure, because a decline can be routed to a human and a confident error cannot.

Where a field's answer set is closed, the item publishes the options (`answer_options`) and the response schema enforces them as an enum (for the three October models, the prompt only). The prompt lists the options in alphabetical order, not in the order they were written, because the key was usually written first. Chance scoring treats such a field as 1/N guessable. A wrong option offered beside the key is a *distractor*. Of 228 answer fields, 85 are booleans, 103 are numbers, 14 are closed sets with published options, and 26, on 24 items, are free text. On those 24 items, the mechanical normaliser of section 3.4 would turn 20 of 458 current wrong answers right, so the remaining wording penalty is small.

### 2.5 Statistics

Scores are reported with Wilson 95% intervals [1]. The intervals treat the items as draws from the set of precise, checkable Australian rules this author would write, not from clinical practice (limitation 4). They do not include run-to-run variation in the models' answers (limitation 13). Two models are called separated when their intervals do not overlap. This rule was set before the first run. As a check, each pair of models was also compared with an exact McNemar test on the items both scored blind, with a decline counted as wrong as in Table 2, and a Holm correction across all 378 pairs. Where shares are pooled across models and items, intervals come from a bootstrap that resamples items, or items and models, as stated.

A chance-adjusted score is also reported. For each item, the probability that a random guesser scores correct is the product over fields of the per-field guess probability (1/2 for a boolean, 1/N for a closed set, 0 for free text and numbers), and the adjusted score is the mean over items of (correct - chance) / (1 - chance).

Any item that no model on the board answers correctly in either condition is treated as unclear rather than hard, excluded from the scored set, and reported as excluded. This rule is population-dependent. Section 3.6 reports its effect, which is now nil.

### 2.6 Failure decomposition, the four cells of a pair

For each model and each item outside the recall-only and equipoise sets, the blind and rule-supplied outcomes place the pair in one of four cells: correct (right in both), *recall failure* (wrong blind, right with the rule), *reasoning failure* (wrong in both) and *distracted* (right blind, wrong with the rule). A pair missing either condition is dropped from the decomposition rather than defaulted to a failure. A blind decline leaves the decomposition entirely. A recall failure shows the model did not produce the current rule unaided. It does not show whether the model held an older version of the rule or none at all.

### 2.7 Exclusion and disclosure policy

A call that returns nothing (a rate limit, a transport fault, a call that runs past the per-cell time budget) is recorded as an error row. It is excluded from every score and reported under its own heading. It is never scored as a wrong answer. A reply that returns but does not parse is reported separately as a parse failure and excluded from scoring. Section 3.10 gives the effect of scoring parse failures as wrong instead.

### 2.8 Held-out items

Thirty items are public, in the repository. One hundred and twelve are held out. A benchmark whose items are published is absorbed into the next generation of training data, after which it measures memorisation of the test [11]. The method is public. The content is held out. As a check on the public items, pooled blind accuracy was 68.3% on the 30 public items and 72.0% on the held-out items. Per model, the public items scored 3.8 points lower on average (bootstrap interval over models 1.5 to 6.3 points lower), and only 7 of 28 models scored higher on them. This shows the public items are not easier for the models. It is not a test of contamination, which would need a canary string or a published hash of the held-out set (limitation 10).

### 2.9 Stopping rule

The build plan, written on 10 July 2026 before any model was run, fixed one go or no-go question: does the leading model fail Australian clinical policy, or clear it? The rule looks at the leading model's blind accuracy and its Wilson interval, not the point estimate, so that one item cannot decide it. There are three outcomes. If the whole interval lies below the threshold, models measurably fail. If the whole interval lies above it, models clear the bar. If the interval crosses it, the suite cannot tell the two apart, and the rule prescribes more items rather than a result. The harness codes these PROCEED, KILL and UNDERPOWERED, after the go or no-go decision each was tied to (section 9). The plan set the threshold at 90%. It was raised to 95% on 7 September 2026, after the version 1.2 results were known (section 3.9, limitation 6). The reason given was clinical: for a rule that decides a subsidy, a bill or a vaccine, a one-in-ten error rate is too high to call a model fit, and one in twenty is the least demanding bar we would defend. The 95% figure is the author's judgement, not a standard.

### 2.10 Release history

| Release | Date | Change | Models |
|---|---|---|---|
| 1.0 | 12-13 July 2026 | First run, 142 items. One immunisation item retagged from equipoise to settled between runs (section 3.10) | 10 |
| 1.1 | 6-7 September 2026 | Fifteen models added. Two billing items repaired and converted to forced choice | 25 |
| 1.2 | 7 September 2026 | Nine more items converted to forced choice, every model re-run on them | 25 |
| 1.3 | 9 October 2026 | Claude Sonnet 5.5 and Claude Haiku 5.5 added | 27 |
| 1.4 | 13 October 2026 | Claude Opus 5.5 added | 28 |
| 1.4.1 | 23 October 2026 | Acute rheumatic fever key corrected to the current guideline. Its blind answers rescored from the stored replies, its rule-supplied cells unmeasured (section 3.8). No model re-run | 28 |

*Table 3. Releases. Versions 1.1 and 1.2 each changed the scoring of some items, so a score is comparable only with scores from the same version or a later version that only added rows. Versions 1.3 and 1.4 only added rows. Version 1.4.1 rescored one item. Every figure in this report is version 1.4.1 unless the text says otherwise.*

## 3. Results

### 3.1 The board

Table 2 gives blind accuracy, the Wilson interval, the chance-adjusted score, the confident-error rate and the decline rate for all 28 models. It is sorted by blind accuracy because the stopping rule uses it. Sorted by confident error instead, the five API-layer leaders and Claude Opus 5.5 still lead, in the same order as by blind accuracy (Claude Fable 5.1 and Claude Opus 5 tie), because most models decline rarely. Two models break the pattern. Claude Haiku 5.5, on the October route, declines 20.4% of items, so its confident-error rate (24.8%) is lower than that of ten models with higher accuracy, and on the items it answers it is right 68.8% of the time. Its decline rate was measured on a route with an unrecorded prompt and default temperature, so part of it may belong to the route. Kimi K2.6, on the API layer, declines 14.7%, and is right on 72.4% of the items it answers. A product that routes a decline to a human gains from a model like these and loses on accuracy. The "Answered" column in Table 2 gives accuracy on answered items for every model.

| Model | Lab | Run | Blind | 95% CI | Chance-adj. | Confident error | Declines | Answered | n |
|---|---|---|---|---|---|---|---|---|---|
| GPT-6 Astra Pro | OpenAI | Sep | 92.7 | 87.1-96.0 | 92.7 | 6.6 | 0.7 | 93.4 | 137 |
| GPT-6 Astra | OpenAI | Sep | 92.0 | 86.2-95.5 | 91.2 | 7.3 | 0.7 | 92.6 | 137 |
| Claude Fable 5.1 | Anthropic | Sep | 83.9 | 76.9-89.1 | 81.9 | 16.1 | 0.0 | 83.9 | 137 |
| Claude Opus 5 | Anthropic | Sep | 83.2 | 76.1-88.5 | 82.1 | 16.1 | 0.7 | 83.8 | 137 |
| Claude Fable 5 | Anthropic | Jul | 82.2 | 74.9-87.8 | 80.1 | 17.0 | 0.7 | 82.8 | 135 |
| Gemini 3.8 Flash | Google | Sep | 79.6 | 72.0-85.5 | 77.3 | 18.2 | 2.2 | 81.3 | 137 |
| Kimi K3 | Moonshot AI | Sep | 78.1 | 70.5-84.2 | 76.6 | 20.4 | 1.5 | 79.3 | 137 |
| Gemini 3.1 Pro Preview | Google | Jul | 77.9 | 70.3-84.1 | 75.1 | 19.9 | 2.2 | 79.7 | 136 |
| GPT-5.6 Terra | OpenAI | Jul | 77.2 | 69.5-83.5 | 75.0 | 22.1 | 0.7 | 77.8 | 136 |
| Grok 4.5 | xAI | Jul | 77.2 | 69.5-83.5 | 74.6 | 19.1 | 3.7 | 80.2 | 136 |
| Grok 4.6 | xAI | Sep | 76.6 | 68.9-82.9 | 72.8 | 19.0 | 4.4 | 80.2 | 137 |
| GPT-5.6 Luna | OpenAI | Sep | 73.7 | 65.8-80.4 | 71.2 | 26.3 | 0.0 | 73.7 | 137 |
| Claude Opus 4.8 | Anthropic | Jul | 72.8 | 64.8-79.6 | 69.6 | 25.0 | 2.2 | 74.4 | 136 |
| Claude Sonnet 5 | Anthropic | Sep | 70.1 | 61.9-77.1 | 67.0 | 26.3 | 3.6 | 72.7 | 137 |
| DeepSeek V4 Pro 0813 | DeepSeek | Sep | 69.3 | 61.2-76.4 | 67.1 | 29.9 | 0.7 | 69.9 | 137 |
| Gemini 3.5 Flash Lite | Google | Sep | 66.4 | 58.2-73.8 | 63.1 | 32.8 | 0.7 | 66.9 | 137 |
| Qwen 3.7 Max | Alibaba | Jul | 64.7 | 56.4-72.2 | 61.8 | 32.4 | 2.9 | 66.7 | 136 |
| DeepSeek V4 Pro | DeepSeek | Jul | 63.2 | 54.9-70.9 | 60.3 | 30.1 | 6.6 | 67.7 | 136 |
| Kimi K2.6 | Moonshot AI | Jul | 61.8 | 53.4-69.5 | 59.2 | 23.5 | 14.7 | 72.4 | 136 |
| DeepSeek V4 Flash | DeepSeek | Sep | 60.6 | 52.2-68.4 | 57.4 | 37.2 | 2.2 | 61.9 | 137 |
| Mistral Large | Mistral | Jul | 59.6 | 51.2-67.4 | 56.5 | 40.4 | 0.0 | 59.6 | 136 |
| Qwen 3.8 Max | Alibaba | Sep | 57.4 | 49.0-65.4 | 52.4 | 36.8 | 5.9 | 60.9 | 136 |
| Qwen 3.8 Flash | Alibaba | Sep | 53.3 | 44.9-61.5 | 48.3 | 45.9 | 0.7 | 53.7 | 135 |
| Mistral Small | Mistral | Sep | 53.3 | 45.0-61.4 | 50.6 | 46.7 | 0.0 | 53.3 | 137 |
| Llama 4 Maverick | Meta | Jul | 51.5 | 43.1-59.7 | 45.6 | 47.8 | 0.7 | 51.9 | 136 |
| *October route* | | | | | | | | | |
| Claude Opus 5.5 | Anthropic | Oct* | 86.1 | 79.4-90.9 | 83.7 | 12.4 | 1.5 | 87.4 | 137 |
| Claude Sonnet 5.5 | Anthropic | Oct* | 74.3 | 66.3-80.9 | 70.5 | 23.5 | 2.2 | 75.9 | 136 |
| Claude Haiku 5.5 | Anthropic | Oct* | 54.7 | 46.4-62.8 | 50.3 | 24.8 | 20.4 | 68.8 | 137 |

*Table 2. All figures are percentages except n. Blind = the question alone, no sources supplied. Answered = accuracy on the items the model did not decline. API-layer rows sorted by blind accuracy, then the October-route rows. The sort is not a ranking (section 3.2). Figures are version 1.4.1 and are not comparable with version 1.0 or 1.1. Section 3.10 accounts for every n below 137. * Run through a Claude Code subscription, not the API layer (section 2.3).*

### 3.2 Separation

By interval overlap, 155 of the 378 model pairs are separated. By paired McNemar tests with a Holm correction, 154 are. Both tests give the same picture at the top. Among the API-layer models, no pair of the five highest scorers, from GPT-6 Astra Pro (92.7%) to Claude Fable 5 (82.2%), is separated, so no order among them is supported. GPT-6 Astra Pro and GPT-6 Astra, the same lab's models at the same list price, are not separated. Gemini 3.8 Flash, next, is separated from the top two under both tests and from none of the other three. Claude Opus 5.5, from the October route, scores 86.1% and is not separated from any of the five. Figure 1 shows the board with each interval, with the October models starred.

FIGURE 1 · BLIND ACCURACY FOR ALL 28 MODELS, WITH THE AUTHOR'S 95% THRESHOLD

## 5 API-SERVED MODELS AT THE TOP: NO PAIR SEPARATED BY INTERVAL OVERLAP OR BY PAIRED TEST

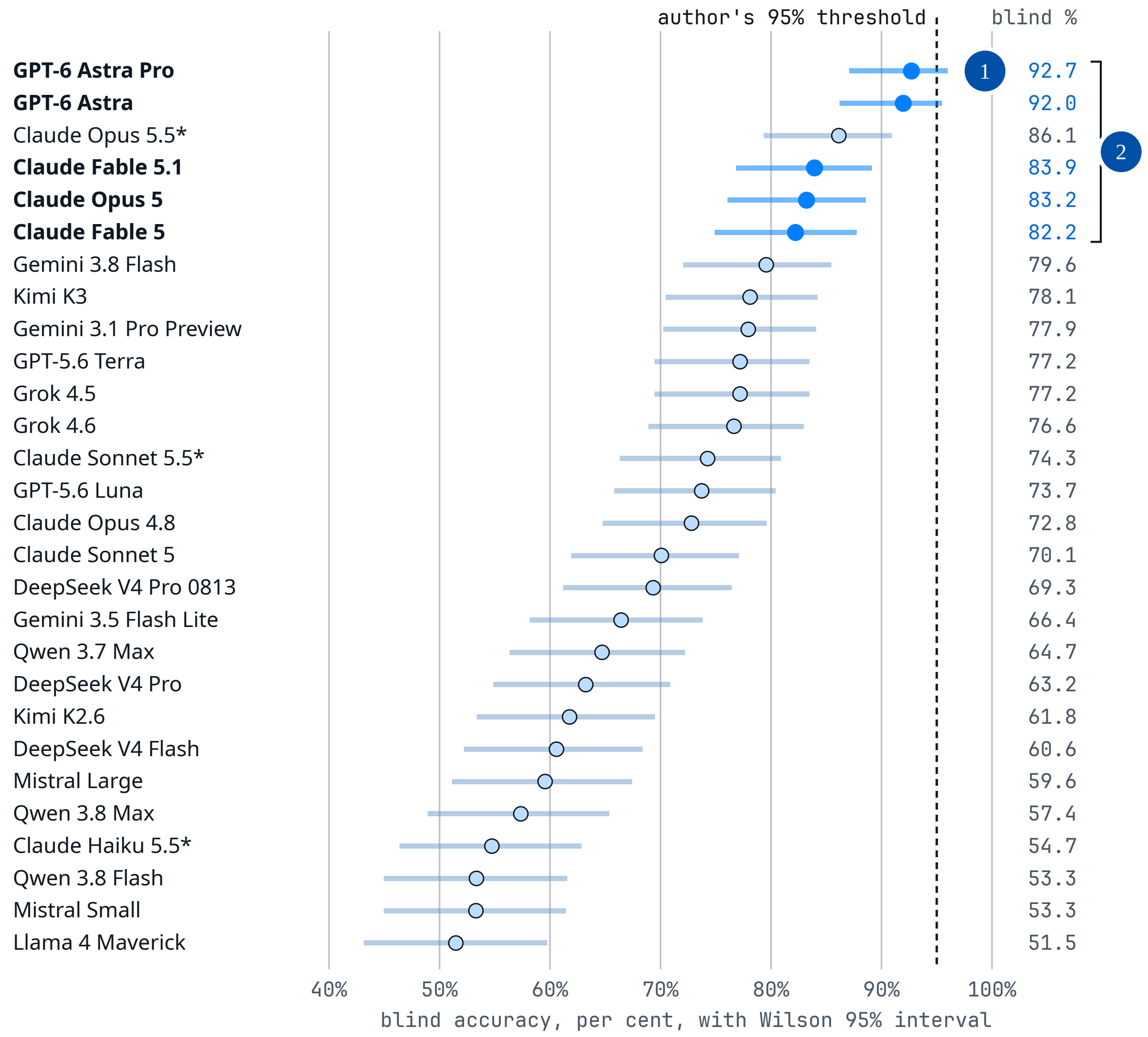

1. *GPT-6 Astra Pro scores 92.7%, interval 87.1 to 96.0. Because the interval crosses the 95% line, the stopping rule cannot decide whether models fail or pass, and asks for more items (sections 2.9 and 3.9).*
2. *Filled blue dots: the five highest-scoring API-served models. No pair of them is separated by interval overlap or by McNemar test with Holm correction, so no order among them is supported. Gemini 3.8 Flash, next, is separated from the top two under both tests. Claude Opus 5.5*, unfilled because it is not API-served, sits among them and is not separated from any of the five (section 3.2). *October route: Claude Code subscription, no schema or temperature control (Table A1).*

Source: AusClinBench v1.4.1 run outputs, re-derived with the benchmark's scorer · 28 models · July to October 2026 · 135 to 137 scored items per model

### 3.3 Price

Prices are OpenRouter list prices in US dollars per million tokens, recorded for the September models on 7 September 2026. The record keeps one list price per model and does not say whether it is the input or the output rate, so the comparison below is coarse, and no price is given for the July or October models.

Gemini 3.8 Flash lists at about $0.75 and scored 79.6%. Claude Fable 5.1 and Claude Opus 5 list at $5 to $10. Gemini 3.8 Flash is not separated from either of them, nor from Claude Opus 5.5 or Claude Fable 5. It is separated from GPT-6 Astra Pro and GPT-6 Astra, which list at about $10, and from eleven models in total, including Qwen 3.8 Flash (53.3%) and Mistral Small (53.3%), which list at about $0.15. Among models listed under $1, blind accuracy spans at least 26 points. That spread is about six times the gap between Gemini 3.8 Flash and the best API-layer model it cannot be told apart from (4.3 points, to Claude Fable 5.1).

In version 1.1 Gemini 3.8 Flash was separated from none of the higher-priced models. In version 1.2 the top two scores rose and their intervals narrowed, which separated them from it.

### 3.4 Failure anatomy, and a scoring artefact we found and removed

The version 1.1 board placed 168 of 3,042 model-item pairs (5.5%) in the reasoning cell, that is, wrong under both conditions. We inspected the items behind most of those pairs by hand. Models had often given the correct answer in different words: "the entire limb" against a key of `entire_limb`, "only if clinically necessary" against `clinically necessary`, "2-6" written with an en dash against a key written with a hyphen. Exact-match scoring cannot tell that from a wrong answer.

Eleven items had free-text keys of this kind. Two were repaired and converted on 6 September (version 1.1) and nine on 7 September (version 1.2). Each was given an explicit option set in the prompt and the response schema, and every model was re-run on it. Scoring remained exact match. Distractors were drawn from values models had actually produced in earlier runs, or from genuine alternatives within the same rule, so that the items would not become trivial. Because a bounded field is guessable, chance rates on the converted fields rose from zero to between 1.3% and 25%, which lowers the chance-adjusted scores.

After the 7 September conversion, reasoning failures fell from 168 pairs to 77. The pair total rose by 30, to 3,072: one item that left the unclear list added 25 pairs (the other was recall-only, outside the decomposition), and changes in blind declines and parse failures on the re-run items added a net 5. The nine converted items went from 107 reasoning pairs to 16. The other items were not re-run, so their 61 stayed 61 by construction. How much of the fall was the wording penalty and how much the easier task of recognising an answer in a list cannot be separated from these data. A mechanical normaliser (case, underscores, dashes and leading articles) recovers 25 of the 107 pairs. Most of the rest differ by paraphrase, which only a human can judge. Every model's blind score rose by between 0.3 and 4.8 points. Version 1.1 and version 1.2 scores are therefore not comparable.

On the version 1.4.1 board the decomposition across 3,398 pairs is: 76.7% correct, 20.1% recall failure (684 pairs), 2.4% reasoning failure (81), 0.8% distracted (26). Recall failures are 86.5% of the 791 failed pairs. A bootstrap that resamples both items and models gives 78.3% to 93.6%. Without the three October models the share is 86.1% (622 of 722). Some supplied clauses state a key value word for word. Of the 119 non-boolean key fields on the 126 items in the decomposition, 48, on 37 items, appear in the clause and not in the blind prompt. Boolean fields were not measured this way, and a clause can make a yes or no answer plain without stating it. On those items the share is 92%. On the others it is 81%, so most of the finding does not rest on the clause stating a non-boolean answer. Per model the share runs from 72% to 100%. The two models at 100% have only 4 and 6 failed pairs. Figure 2 shows the split and the gain for each model when the rule is supplied.

Two converted items show that the forced-choice form still separates a right answer from a wrong one. One, which turns on a clause allowing exceptions, still fails: 11 of 28 models answer it correctly blind and only 15 of 28 with the governing clause supplied. Every one of the 13 wrong answers given with the clause selects "no exception applies", against a clause reading "unless there are exceptional circumstances". Another, a PBS authority-code item tagged recall-only, remains at 0 of 28 blind and 2 of 27 with the clause, with blind answers spread across all four distractors. A third converted item moved from 0% to 96% blind (27 of 28 models). That fits both explanations above: the models knew the answer and had worded it differently, or the list made it easier.

FIGURE 2 · SUPPLY THE RULE, AND MOST FAILURES GO

# 86% OF FAILED PAIRS WERE RECALL FAILURES: WRONG BLIND, RIGHT ONCE THE RULE WAS SUPPLIED

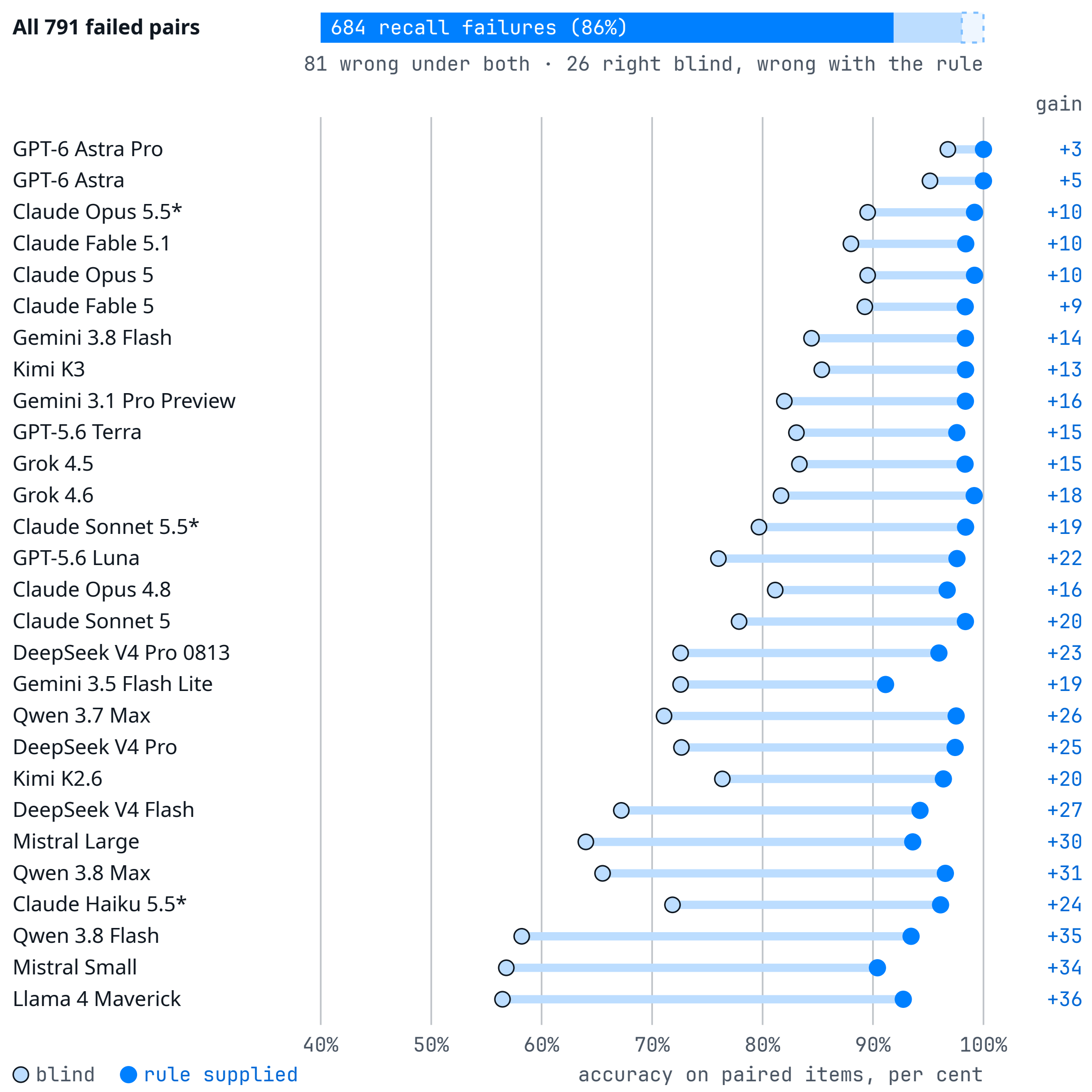

1. *Open dot: blind, the question alone. Filled dot: the same question with the governing clause supplied. Gain: rule-supplied accuracy minus blind accuracy, in points. *October route: Claude Code subscription, no schema or temperature control (Table A1).*
2. *Paired items only: 3,398 model-item pairs, with recall-only items, equipoise items and blind declines removed (section 2.6). Removing them puts every blind figure 2 to 17 points above Table 2, most for Claude Haiku 5.5 and Kimi K2.6, which decline often. The rheumatic fever item has no rule-supplied run in this release (section 3.8). Rows are ranked by blind accuracy as in Table 2, with the October models placed among the rest. Wrong under both conditions: 81 pairs. Right blind, wrong with the rule: 26. A bootstrap over items and models puts the recall-failure share at 78.3% to 93.6%. Without the three October models it is 86.1% (section 3.4).*

Source: AusClinBench v1.4.1 run outputs, re-derived with the benchmark's scorer · 28 models · July to October 2026

### 3.5 Distraction and equipoise

Twenty-six pairs (0.8%) were answered correctly blind and incorrectly once the correct clause was supplied. It is the only cell that measures whether supplying context degrades a model, which is the condition a retrieval-equipped product runs in. But two identical July runs of one model disagreed on 2.8% of cells (section 3.11). The distracted cell (0.8%) and the reasoning cell (2.4%) both sit inside that rate, so neither can be read as an effect of the model alone. The recall-failure cell (20.1%) is far above it. The 2.8% comes from one model at temperature 0 under version 1.0 scoring, and does not describe the October rows, which ran at default temperature.

The equipoise items ask about a point on which the guideline declines to give a directive, so the correct answer is to say the evidence is insufficient. All five equipoise items are immunisation items. Across them, 115 of 140 blind answers (82.1%) gave a directive the guideline does not give. A model scores an equipoise pair only when it gives no directive on the equipoise item and commits on its settled partner. Only 23 of 140 pairs (16.4%) were right on both. With five answers per model, no model's rate is precise enough to rank, so only pooled figures are reported.

### 3.6 Effect of the population-dependent exclusion rule

Adding fifteen models in version 1.1 moved the ten July scores by up to about 1.4 points. That figure mixes two causes, the population effect described here and the two-item repair made at the same time, and the two cannot be separated in the stored files. Items that no July model had answered in either condition had been excluded from the July board as unclear. Some returned to the scored set once a September model answered them, and then counted against every July model. The July board is not a defined release in the harness, so this figure is approximate.

Since version 1.2 the exclusion list has been empty: every one of the 142 items is answered by at least one model in at least one condition. The rule is inactive and is kept only so that later boards run the same code.

### 3.7 Domain

Pooled blind accuracy across all 28 models, equipoise items excluded: PBS authority 56.0% (671 blind answers, 24 items), MBS billing 64.8% (1,453, 52 items), Ahpra obligations 77.7% (364, 13 items), triage red flags 79.3% (560, 20 items), immunisation and screening 87.6% (773, 28 scored items). Answers to the same item are not independent, so Figure 3 gives intervals from a bootstrap over items, and the differences between domains were bootstrapped the same way. PBS authority is below triage red flags (difference -40.3 to -6.2 points), Ahpra obligations (-39.8 to -2.9) and immunisation and screening (-46.6 to -15.7). MBS billing is below triage red flags (-27.0 to -1.7) and immunisation and screening (-33.8 to -11.1). PBS authority is not separated from MBS billing (difference -24.9 to +6.8 points), and no other ordering is supported. Ahpra and triage rest on 13 and 20 items. The leading model scores 79.2% on PBS authority and 92.3% to 100% on the other four domains, a domain average of 93.6% against its pooled 92.7%.

FIGURE 3 · ACCURACY BY DOMAIN

## 56% IN PBS AUTHORITY, FROM ONLY 24 ITEMS: BELOW THREE OF THE OTHER FOUR DOMAINS

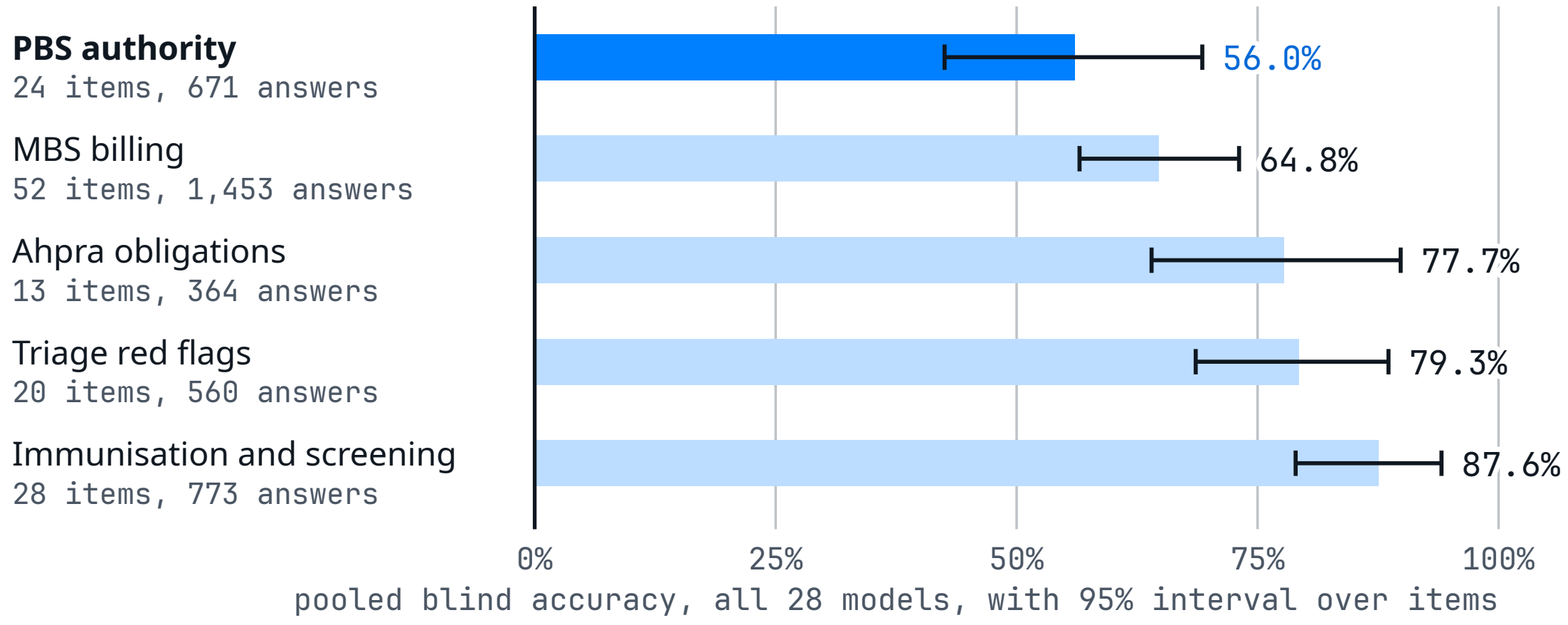

1. *Bars: blind answers pooled across the 28 models. Whiskers: 95% interval from a bootstrap that resamples items, because answers to the same item are not independent.*
2. *A bootstrap of the differences over items puts PBS authority below triage, Ahpra and immunisation, and MBS billing below triage and immunisation. PBS authority is not separated from MBS billing, and no other order is supported (section 3.7).*

Source: AusClinBench v1.4.1 run outputs, re-derived with the benchmark's scorer · 28 models · July to October 2026 · equipoise items excluded

### 3.8 Aboriginal and Torres Strait Islander health

Twenty items carry a First Nations tag, set when each item was written. On these items the answer depends on the patient being Aboriginal or Torres Strait Islander (ATSI), or the rule comes from guidance written for Aboriginal and Torres Strait Islander people. They sit in four domains (seven immunisation, five triage, four MBS, four PBS), and eight are public.

Overall, the models did no worse on ATSI items. Pooled blind accuracy was 79.5% on the 20 tagged items (560 answers) and 69.8% on the other 117. The tagged items sit mostly in the easier domains, so we compared them with untagged items in the same domain. Within domains, the tagged items scored 4.7 points higher, with a bootstrap interval over items of -7.5 to +15.2, so no difference is shown in either direction. With the rule supplied, accuracy on the tagged items was 97.9%, and 88.5% of their failed pairs were recall failures (86.2% on the other items). The confident-error rate on the tagged items was 18.2%, against 27.1% on the rest.

The wrong answers on these items fall into three kinds: the rule for the general population, an older Australian rule, or a rule from another country. Each kind rests on a few items, so these are examples, not rates.

- **Side-by-side items.** Four items ask the same question for a non-Indigenous patient and an Aboriginal or Torres Strait Islander patient. Across 112 answers, 13 got the Aboriginal or Torres Strait Islander half wrong and 6 got the non-Indigenous half wrong. In 11 answers only the Aboriginal or Torres Strait Islander half was wrong, and in 4 only the non-Indigenous half was (exact test $p = 0.12$, so not significant). In 6 of the 13, the model gave the value that applies to the general population.
- **A recently changed vaccine recommendation.** On a held-out immunisation item about an adult recommendation for Aboriginal and Torres Strait Islander people, all 28 models were wrong blind. Twenty-five gave the earlier

Australian rule. With the clause supplied, all 28 were right. This one item explains most of the gap in immunisation, where tagged items scored 75.0% against 91.9% for the rest (difference -16.9 points, interval -43.7 to +1.7). Without it, the tagged immunisation items scored 87.5%.

- **Otitis media.** On a held-out item from the Australian otitis media guideline for Aboriginal and Torres Strait Islander children, 8 of 28 models gave the antibiotic dose recommended in guidance from another country. With the guideline supplied, 27 of 28 were right.
- **Acute rheumatic fever: a wrong key, now corrected.** A public item asks for the minimum duration of secondary prophylaxis for an Aboriginal teenager after one episode of acute rheumatic fever without heart disease. Release 1.4 keyed it 10 years or until age 21, from a 2017 review [12]. The Australian guideline changed this to a minimum of 5 years after the last episode or until age 21, whichever is longer, in its 2020 edition, and the current edition (3.4, October 2026) keeps it [13]. So 5 years was the rule on every run date. Under the old key, 19 of 28 models gave the current answer and were marked wrong, and 8 gave the 2017 answer and were marked right. Release 1.4.1 corrects the key and rescores the stored blind answers: 18 models are now right, and one more gave 5 years but also declined. The rule-supplied condition had supplied the 2017 text, so the item's rule-supplied cells are unmeasured (section 3.10). The correction moves every model by 0.7 points, up or down, and changes no finding. It does change how fragile the stopping rule is (section 3.9).

In the vaccine and otitis media items, the models were wrong blind and right once the rule was supplied. What they lacked was the rule that applies to Aboriginal and Torres Strait Islander patients where it differs from the default. A wrong answer of that kind could widen an existing gap in care, though this study did not measure harm. The rheumatic fever item shows the same risk in the answer key: a key written from an older source marked the current rule wrong. Twenty items are too few to estimate the size of any difference.

### 3.9 Stopping rule

Section 2.9 defines the rule. Raising the threshold from 90% to 95% after the version 1.2 results were known made "models fail" easier to reach and "models pass" harder (limitation 6).

In version 1.0 the leading model's entire interval lay below the threshold, and the verdict was that frontier models measurably failed. That finding no longer stands. In version 1.4.1 GPT-6 Astra Pro scores 92.7% (127 of 137) with an interval of 87.1 to 96.0. The interval crosses 95%, and it also crosses the original 90%, so the verdict is undecided under either threshold. At 142 items the suite cannot tell a model that clears the bar from one that fails one item in seven. If GPT-6 Astra Pro kept scoring 92.7%, about 350 items would be needed to put its whole interval below 95%. In version 1.4, before the key correction, the figure was about 205.

The verdict depends on the keys, in both directions. With 127 correct, one wrong key among the leader's correct answers would leave its interval (86.2 to 95.5) still crossing 95%. Two would put it wholly below (85.3 to 94.9), and the verdict would become "models fail". In the other direction, "models pass" needs 136 or 137 of 137 correct. In version 1.4, with the rheumatic fever key wrong (section 3.8), the leader had 126 correct, and a single further wrong key would have given "models fail". The leader is also the highest of 28 noisy estimates, so its score is more likely to be high by chance than any other model's. The rule as written does not correct for that, and it has no pre-set maximum number of items, so repeated looks as items are added would inflate its error.

The coded version of the rule in the harness gives no verdict here, because it refuses to rule while any cell is unmeasured, and 38 are (section 3.10). The verdict above is the same rule applied by hand to the leading model's interval.

### 3.10 Unmeasured cells and the n in Table 2

Thirty-eight of 7,952 cells have no scorable result and are excluded rather than counted as errors. Twenty-eight are the rule-supplied cells of the rheumatic fever item, which carried the superseded clause and are unmeasured in version 1.4.1 (section 3.8). Of the other ten, four are unparseable replies from Claude Fable 5 in July, and two are unparseable replies from Claude Sonnet 5.5 in October (each reply gave two JSON objects, the second a self-correction). These are real model behaviour, so they are reported, not re-rolled. Three are provider timeouts cut off by the per-cell deadline during the conversion re-run, and one is an earlier provider rate-limit rejection (HTTP 429). Scoring the three blind parse failures as wrong would move Claude Fable 5 from 82.2% to 81.0% and Claude Sonnet 5.5 from 74.3% to 73.7%.

The blind n in Table 2 is 137 (142 less the five equipoise items) less two things. Six blind cells are among the ten above. The rheumatic fever correction rescored blind cells and removed none. One immunisation item was tagged equipoise when nine July models ran and was retagged as a settled item before September. The scorer reads the tag from each run's record, so that item is off the blind score for those nine models, which is why their n is 136. Claude Fable 5, also a July model, has rows that carry the settled tag, so it was scored on the item. All nine gave the answer the current key accepts, so scoring them on it would raise each by 0.2 to 0.4 points. No separation and no verdict would change. Table 2 keeps the benchmark scorer's figures so that it can be re-derived. Together these account for all 15 missing blind cells. Two further models are excluded from the board entirely: Gemma 4 31B, for which every call was rejected, and Qwen 3.8 27B, which returned 67 empty replies of 284.

### 3.11 July-only experiments (version 1.0, ten models)

These ran in July 2026, under version 1.0 scoring, and were not repeated. Their figures are not comparable with the rest of this report. The web-search, repeat-run and wrong-answer figures were re-derived from the July run files. The wrong-answer classification was called "severity" in version 1.0. It classifies the type of wrong answer, not the harm it would do, and no harm rating has been made. The pushback and rewording figures come from scripts outside the benchmark harness and cannot be reproduced from it.

- **Web search.** With live web search enabled and no clause supplied, every model improved, to between 84% and 95% accuracy, including on items built so that the most findable answer online is a superseded or foreign rule. The expectation written in the build plan before the run, that search would hurt on those items, was not supported.
- **Pushback** (not reproducible from the harness). Under a one-line challenge to a correct answer, most models kept their answer. One model gave way on 20% of challenged items.
- **Rewording** (not reproducible from the harness). Across the ten models, answer consistency under paraphrase ranged from 67% to 93% (n=15 items).
- **Repeat run.** Two identical runs of one model agreed on 97.2% of cells (276 of 284).
- **Type of wrong answer.** Scored against the version 1.4.1 keys, the July run gave 571 blind wrong answers. Their distinct wrong values were researched against primary sources, and 153 answers could be classified: 66% gave a superseded Australian rule (stale), 22% a rule from another country, and 12% a value that research confirmed no rule contains (fabricated). The other 418 could not be classified. The July labels had the rheumatic fever item the wrong way round, counting six correct answers as a foreign rule (section 3.8). This is the only direct evidence in this report that recall failures include out-of-date rules.

## 4. Discussion

AI models now sit inside the software Australian clinicians use, and that is unlikely to reverse.

This benchmark shows a wide spread in how well models know Australian clinical rules, from 51.5% to 92.7% blind. For safety, the rate of confident wrong answers is as important as accuracy, because a model that says it does not know, as an experienced clinician would when there is no clear answer, leaves the decision with a person with expert judgement. Confident-error rates ranged from 6.6% to 47.8%. These rates carry no intervals here, and nobody checked whether each decline was appropriate, so they show a spread between models rather than a ranking. Even so, a benchmark like this can start to compare the safety of the models inside clinical products.

On this item set, most failures are failures to produce the current rule unaided. Handed the exact clause, every model on the board scored at least 90% on the paired items, and recall failures are 86% of failed pairs. The July web-search result points the same way: given a way to look the rule up, every model improved. The design shows that the rule was not recalled.

For a vendor, the result says that what reaches the model at the moment of the question matters. It does not say that retrieval makes a product safe. The clause in this study was the exact one the key was written from. A product that retrieves a superseded clause, a neighbouring clause or nothing was not tested. Even with the clause supplied, the best models made no errors and the worst got 9.6% wrong. For a practice, the useful question to ask a vendor is where its rules come from and how current they are, as well as which model it uses. The products most likely to meet these questions unprompted are inbox and decision-support assistants that draft billing or prescribing advice. A transcription tool meets them only when it suggests an item number or a script.

Three findings deserve more weight than their size suggests. On five immunisation items, models gave a directive on 82% of answers where the guideline gives none. Two models decline often enough that their confident-error rate is low despite low accuracy, which a product with a human fallback could use, though one of them (Claude Haiku 5.5) was measured on the October route. And PBS authority, the lowest-scoring domain, is one where a wrong answer costs a patient a subsidy. That ranking of stakes is the author's judgement. Triage red flags, where a wrong answer can cause harm, sit in the middle. None of these figures has a clinician baseline, so 56% on PBS authority cannot yet be called worse or better than practice. In the author's clinical experience, prescribing software supplies most PBS authority detail, so the risk is an assistant that contradicts it.

The Aboriginal and Torres Strait Islander items (section 3.8) did not score lower overall. On a few of them, though, the wrong answers gave the rule for the general population, an older Australian rule or another country's rule. On these items, an error of that kind would deny a patient a funded vaccine or give a child the wrong dose. A benchmark written for another country cannot test for it.

This benchmark is not evidence about the performance or regulatory status of any product built on a model, and it is not a clinical validation. It can still inform how to regulate a technology that moves fast. In Australia, software intended for clinical use may be regulated as a medical device by the Therapeutic Goods Administration (TGA). We propose treating the AI model as an 'ingredient' of the end product, with a safety measure of its own. The TGA already does something similar for listed medicines, which may contain only ingredients on a permitted list [14]. A benchmark like this could set the standard a model must meet before a product can include it, for example a minimum accuracy and a maximum confident-error rate on local rules. Manufacturers would keep their choice of model within that standard.

The results also show the limits of that proposal. A model's score changed by up to 36 points depending on whether the rule was supplied, so the product around the model decides much of its safety. A model standard would be a minimum for the ingredient. It would not replace evaluation of the product, just as a permitted ingredient does not make a medicine safe. It would also need things this study does not have: answer keys reviewed by independent clinicians, a body that holds the held-out items and re-tests models as the rules change, and a threshold set before results are known. No model on this board has been shown to meet a 95% standard. The proposal accepts that the technology is already in use. It puts a broad safety requirement on the model, a general-purpose component, while research works out how to make these products safe in the long term.

## 5. Limitations

1. **Snapshot.** The July, September and October runs measure the models as served on those dates and the rules current then. Keys were checked when written and by the freshness check, not re-verified on each run date. The freshness check did not catch the rheumatic fever key, which was out of date before the first run (section 3.8). The prompt names Australia but gives no date. A later run under the same model name is not the same measurement.

2. **Exact match is harsh, and in eleven items it was wrong.** Section 3.4 measures the effect and describes the correction. Twenty-four items still have a free-text field, so the reasoning cell should be read as an upper bound.
3. **Single author.** All keys were written by one general practitioner. Independent review of the keys by two RACGP fellows is planned and has not happened. Until it has, the keys are one clinician's reading of the source, and two wrong keys among the leader's correct answers would change the verdict (section 3.9). A pharmacist for PBS authority and a billing specialist for MBS billing would add to that review. One July key was found wrong and corrected in the models' favour. A second, on acute rheumatic fever, was written from a 2017 source that the 2020 guideline replaced. It was found during the analysis in section 3.8 and is corrected in version 1.4.1. Model consensus is no evidence about a key, in either direction: on one billing item eight of nine models agreed with each other and against the primary source, and on the rheumatic fever item 19 of 28 agreed with the current guideline against the key.
4. **The item set is not yet a generalisable sample of practice.** It over-samples rules that are precise, checkable and easy to get wrong, and MBS billing is 37% of it.
5. **Australian policy only.** Nothing generalises to another jurisdiction. Variation between states was not tested.
6. **Power and the stopping rule.** With 142 items many comparisons are not significant: 155 of 378 pairs separate by overlap and 154 by paired tests. The stopping rule returns undecided (section 3.9). Its threshold was raised from 90% to 95% after the version 1.2 results were known. The verdict is undecided at either value, so the change does not manufacture a conclusion, but a reader should know the sequence. The rule tests the leading model, which is chosen after the fact, and has no pre-set maximum size.
7. **Population-dependent exclusion.** Section 3.6. Scores are comparable within a board, not across boards of different composition, and version 1.1 and 1.2 scores are not comparable with each other because the scoring changed between them (section 3.4).
8. **Not a product evaluation.** The benchmark measures the base model's policy localisation. It does not evaluate any product built on a model, and it does not license any model for clinical use. No human clinician baseline has been measured.
9. **Two serving routes.** The three October models ran through a Claude Code subscription, with the answer format requested in the prompt rather than enforced, no temperature set, no data-collection control, and the route's own prompt, tools and thinking setting unrecorded (section 2.3). Their rows are a separate cohort until they are re-run through the API layer. No model on either route had a reasoning setting recorded.
10. **Held-out items on an uncontrolled route.** All 142 items, including the 112 held out, were sent through the October route without data-collection protection. Their held-out status cannot be guaranteed against future training by that developer. A canary string in the public items and a published hash of the held-out set would let later contamination be checked.
11. **The rule-supplied condition is the best case.** The supplied clause is the one the key was written from, and on 37 items it states a key value word for word (section 3.4). No wrong-clause or superseded-clause condition was run.
12. **Distractors were harvested from earlier runs.** On the converted items, some distractors are answers the July and September models gave, so those items are tuned to the board that produced them. The October models, whose answers were never harvested, found the converted items about as much harder than the rest (5.5 points, on 33 answers) as the other models did (3.7 points), so there is no sign the harvest made them harder for the models it came from. On 33 answers this comparison cannot detect a modest effect.
13. **Run-to-run noise.** Only one repeat run of one model exists (section 3.11). The intervals in this report do not include run-to-run variation, and the October rows ran at default temperature.

## 6. Conclusion

AI models already inform Australian clinical work, and they differ widely in how well they know the rules that govern it. Blind accuracy ran from 51.5% to 92.7%, and confident errors from 6.6% to 47.8%. Most errors were failures to recall the current rule, and most disappeared once the rule was supplied. On a few Aboriginal and Torres Strait Islander health items, models gave the rule for the general population, an older rule or another country's rule. We propose that regulators treat the model as an 'ingredient' of a clinical product, with a minimum standard on local rules that any model in the product must meet. That standard would sit beside product evaluation, not replace it, and it needs answer keys reviewed by independent clinicians and refreshed as the rules change. We will continue to update and improve our benchmark. The next release will add that review, more items, and a re-run of the three October models through the API layer.

## 7. Data and code availability

The leaderboard, methods and changelog are public at ausclinbench.vercel.app. The site shows the current release, which may include models added after this report (release 1.5.1 adds Gemini 4 Argon to the 28 models here). The harness, the scorer, the test suite and the 30 public items are available from the author on request, under MIT for code and CC BY 4.0 for items. The 112 held-out items are not published. Raw run outputs contain held-out item identifiers and model output and stay local. Every published figure re-derives from the stored run outputs with the benchmark's scorer. The system prompt and the per-model serving details are in Appendix A.

## 8. Ethics

No patient data and no human participants were involved at any stage, so no human research ethics review was sought. The planned independent key review and any clinician baseline would involve people, and the baseline would need ethics review before it starts. Items are written from published rules. Provider-side data collection was disabled on every call through the API layer. The three October models ran through a Claude Code subscription, which has no per-call data-collection control, so the held-out items sent on that route carried no such protection (limitation 10).

## 9. Funding and competing interests

**Funding.** None. The benchmark was designed, authored and run by the author without external funding. No grant, institution, vendor or client funded any part of it, and no organisation had any role in the design, the item writing, the analysis, or the decision to publish.

**Competing interests.** The author is a practising Australian general practitioner, and the items are written from the rules he works under. He is also founder and director of The Long Game Project, a scenario-exercise and wargaming consultancy. Advisory work on clinical AI assurance, which that consultancy could offer, is a plausible commercial beneficiary of this benchmark's visibility. The stopping rule ties the project's next step to its result: a "models fail" verdict would continue the project, and raising the threshold made that verdict easier to reach. Other than paid access to the models, described next, the author holds no equity, consultancy, advisory role, grant or other financial relationship with any of the model vendors ranked here, and none was contacted before publication. Twenty-five models were accessed at public list prices through a commercial API aggregator, which pays the vendors. The three October models were accessed through a paid Claude Code subscription from Anthropic. The scoring is deterministic, and every published figure re-derives from the benchmark's scorer and the stored run outputs. A reader can request the code and the 30 public items to check the method. The held-out items and raw outputs are not shared (section 7), so the remaining figures rest on the code and this report.

## References

1. Wilson EB. Probable inference, the law of succession, and statistical inference. Journal of the American Statistical Association. 1927;22(158):209-212.

2. Australian Government Department of Health. Schedule of Pharmaceutical Benefits. https://www.pbs.gov.au/
3. Australian Government Department of Health. MBS Online. http://www.mbsonline.gov.au/
4. Australian Technical Advisory Group on Immunisation. Australian Immunisation Handbook. https://immunisationhandbook.health.gov.au/
5. Australian Health Practitioner Regulation Agency and Medical Board of Australia. Codes, guidelines and policies. https://www.ahpra.gov.au/
6. Jin D, Pan E, Oufattole N, et al. What disease does this patient have? A large-scale open domain question answering dataset from medical exams. Applied Sciences. 2021;11(14):6421. doi:10.3390/app11146421
7. Singhal K, Azizi S, Tu T, et al. Large language models encode clinical knowledge. Nature. 2023;620(7972):172-180. doi:10.1038/s41586-023-06291-2
8. Singhal K, Tu T, Gottweis J, et al. Toward expert-level medical question answering with large language models. Nature Medicine. 2025;31(3):943-950. doi:10.1038/s41591-024-03423-7
9. Zheng L, Chiang WL, Sheng Y, et al. Judging LLM-as-a-judge with MT-Bench and Chatbot Arena. Advances in Neural Information Processing Systems 36. 2023:46595-46623. doi:10.52202/075280-2020
10. McIntosh TR, Susnjak T, Arachchilage N, et al. Inadequacies of large language model benchmarks in the era of generative artificial intelligence. IEEE Transactions on Artificial Intelligence. 2026;7(1):22-39. doi:10.1109/tai.2025.3569516
11. Kaneko M, Baldwin T. Investigating how pre-training data leakage affects models' reproduction and detection capabilities. Proceedings of the 2025 Conference on Empirical Methods in Natural Language Processing. 2025:23556-23566. doi:10.18653/v1/2025.emnlp-main.1201
12. Ralph AP, Noonan S, Boardman C, Halkon C, Currie BJ. Prescribing for people with acute rheumatic fever. Australian Prescriber. 2017;40(2):70-75. doi:10.18773/austprescr.2017.011
13. Ralph A, Currie B, Marangou J, eds. Australian guideline for the prevention, diagnosis and management of acute rheumatic fever and rheumatic heart disease (Edition 3.4). Menzies School of Health Research; October 2026. Table 10.3, Recommended duration of secondary prophylaxis. https://rhdaustralia.org.au/arf-rhd-guidelines/
14. Therapeutic Goods (Permissible Ingredients) Determination (No. 1) 2025. Made under section 26BB of the Therapeutic Goods Act 1989. https://www.legislation.gov.au/F2025L00212

## Appendix A. Prompt and serving details

**System prompt, both conditions, every model** (verbatim). On both routes it was sent as the request's system prompt. Whether the October route added a prompt of its own was not recorded.

> You are answering a question about Australian clinical policy: the Pharmaceutical Benefits Scheme, the Medicare Benefits Schedule, Ahpra and Medical Board obligations, immunisation and screening program rules, or urgent-referral red flags. Answer strictly as the Australian rules stand, not as clinical practice in another country. Respond with JSON only, matching the schema. When the current Australian rules genuinely do not let you answer, set the boolean field "abstain" to true instead of guessing; the other fields are then ignored. When you can answer, set "abstain" to false and complete every field.

The user message contains the vignette and question, and in the rule-supplied condition the governing clause. On the October route the JSON schema is appended to the user message as an instruction.

| Model | Route | Providers recorded | Run dates (2026) |
|---|---|---|---|
| GPT-5.6 Terra, Gemini 3.1 Pro Preview, Grok 4.5, Qwen 3.7 Max, Mistral Large | API layer | The model's own lab | 12 Jul, 6-7 Sep |
| Claude Opus 4.8 | API layer | Anthropic, Claude Platform on AWS | 12 Jul, 6-7 Sep |
| Claude Fable 5 | API layer | Anthropic, Amazon Bedrock | 12 Jul, 6-7 Sep |
| DeepSeek V4 Pro | API layer | Baidu, Ionstream, StreamLake | 12 Jul, 6-7 Sep |
| Kimi K2.6 | API layer | Baidu, Decart, Inceptron | 12 Jul, 6-7 Sep |
| Llama 4 Maverick | API layer | DeepInfra, DigitalOcean | 12 Jul, 6-7 Sep |
| GPT-6 Astra Pro, GPT-6 Astra, GPT-5.6 Luna, Gemini 3.8 Flash, Gemini 3.5 Flash Lite, Grok 4.6, Mistral Small, Qwen 3.8 Max, Qwen 3.8 Flash, Claude Fable 5.1 | API layer | The model's own lab | 6-7 Sep |
| Claude Opus 5, Claude Sonnet 5 | API layer | Claude Platform on AWS | 6-7 Sep |
| DeepSeek V4 Pro 0813 | API layer | Alibaba, Baidu | 6-7 Sep |
| DeepSeek V4 Flash | API layer | Baidu, DeepInfra, OpenInference | 6-7 Sep |
| Kimi K3 | API layer | Makora, Sail Research | 6-7 Sep |
| Claude Sonnet 5.5, Claude Haiku 5.5 | Claude Code subscription | Anthropic | 9 Oct |
| Claude Opus 5.5 | Claude Code subscription | Anthropic | 13 Oct |

*Table A1. Serving details. API layer: OpenRouter, temperature 0, schema enforced, fallbacks off, data collection denied. No model on either route was sent a reasoning or effort setting. July models were re-run on 6 and 7 September only on the converted items. A few September calls that errored carry no provider.*

API-layer slugs, as requested: openai/gpt-6-astra-pro, openai/gpt-6-astra, openai/gpt-5.6-terra, openai/gpt-5.6-luna, anthropic/claude-fable-5.1, anthropic/claude-fable-5, anthropic/claude-opus-5, anthropic/claude-opus-4.8, anthropic/claude-sonnet-5, google/gemini-3.8-flash, google/gemini-3.1-pro-preview, google/gemini-3.5-flash-lite, moonshotai/kimi-k3, moonshotai/kimi-k2.6, x-ai/grok-4.6, x-ai/grok-4.5, deepseek/deepseek-v4-pro-0813, deepseek/deepseek-v4-pro, deepseek/deepseek-v4-flash-0731, qwen/qwen3.7-max, qwen/qwen3.8-max-0902, qwen/qwen3.8-flash, mistralai/mistral-large-2512, mistralai/mistral-small-2603, meta-llama/llama-4-maverick. October route: claude-opus-5-5, claude-sonnet-5-5, claude-haiku-5-5.

## Appendix B. A worked item

One public item, ahpra-obligations-0003, as a model sees it in the rule-supplied condition. The blind condition is the same without the first block.

> **Governing authority text:** [1] A person must not advertise a regulated health service ... in a way that-- ... (c) uses testimonials or purported testimonials about the service or business. Maximum penalty-- (a) in the case of an individual--$60,000
>
> **Scenario:** A general practice wants to add a "What our patients say" section to its website, quoting three unsolicited Google reviews in which patients praise the quality of the care they received from a named doctor.

> The reviews are genuine and unpaid.
>
> **Question:** Is this advertising permitted under the National Law, and what is the maximum penalty for an individual?
>
> Return a single JSON object with exactly these keys: abstain, advertising_permitted, maximum_penalty_individual_aud

Key: `advertising_permitted` false, `maximum_penalty_individual_aud` 60000. Both fields must match. Blind, 12 of 28 models were right. With the clause, 28 of 28 were. The item is a clean example of a recall failure, and also of the leakage in limitation 11: the clause states the penalty, so the rule-supplied condition tests reading as well as reasoning.